# The theory of homogeneity of nonlinear structural systems - A general basis for structural safety assessment

Dr.-Ing. habil. Tammam Bakeer, Dresden, Germany

E-Mail: mail@bakeer.de

Date: December 1, 2022

## Abstract

The paper develops a novel and general methodology to characterize the nonlinearity of structural systems and to provide a mathematically proven basis for applying partial safety factors to nonlinear structural systems. It establishes, for the first time since the development of *limit-state theory*, the necessary key relationship between the *partial safety factor concept* and the *reliability theory* of nonlinear structural systems.

The *degree of homogeneity* has been introduced as a nonlinearity measure at the design point, allowing an efficient mathematical decoupling of the *reliability index* into nonlinearity-invariant *partial reliability indexes*. With this formulation, critical safety situations in extreme cases of nonlinearities have been identified in complex nonlinear structural systems.

The theory resulted in two main outcomes based on the asymptotic behaviour of the reliability index. First, the reliability index of any nonlinear structural system remains always bounded between an upper and lower bound which can be determined by the concept of nonlinearity-invariant *partial reliability indexes*. The second is nonlinearity-invariant *critical partial safety factors*, a concept that assures a reliability index greater than the target reliability index in any nonlinear structural system.

Homogeneity analysis has been suggested to assess the safety of complex nonlinear structural systems. While it can be coupled with advanced computational methods available in structural mechanics, it is not specifically designed for engineering practice. The proposed theory is designed primarily to provide code writers with the necessary procedure for calibrating partial safety factors for nonlinear structural systems, and to identify the over-safe or under-safe cases in the codes of practice.

## Keywords

## Abbreviations

| | | | |
|---|---|---|---|
| COV | Coefficient of variation | PDH | Partial degree of homogeneity |
| DH | Degree of homogeneity | PRI | Partial reliability index |
| DHN | Degree of homogeneity of the non-lognormality | PSF | Partial safety factor |
| ECOV | Estimating the coefficient of variation | RI | Reliability Index |
| GSF | Global safety factor | RM | Resistance model |
| LRFD | Load and resistance factor design | RPDH | Relative partial degree of homogeneity |
| LSF | Limit-state function | RSP | Relative sensitivity parameter |
| OP | Over-proportional or over-linear | TRI | Target reliability index |
| | | UP | Under-proportional or under-linear |

## 1  Introduction

Safety factor, as the first proposed measure of structural safety, has provided over centuries an efficient and practical deterministic evaluation of structural safety. It still plays a major role in modern design standards and engineering practice, even with the existence of the most powerful probabilistic methods and computer machines. The term was probably first used by *Forest de Belidor* in his work in 1729 [1–3]. However, the evaluation of safety in terms of statistical properties of the basic variables of the structural system was first introduced by *Max Mayer* in 1926 [4]. The concept is based on uncertainty propagation "*Fehlerfortpflanzung*", which was originally introduced by *Carl Friedrich Gauss* in his work on "*the theory of the combination of observations least subject to error*" [5]. In today's terms, it is known as the variance formula, and it is a direct result of the first-order *Taylor's series expansion* of independent variables. With the Gaussian variance formula, Mayer was aware that the variance in the response of the system could be affected by both the nonlinearity of the system itself and by the statistical properties of its basic variables.

Mayer's views of structural safety remained unknown until the 1950s when repeated criticisms were raised about inefficient designs and inconsistent reliability resulting from the application of safety factors [6]. As a result, the second-moment approach has been developed, and a new measure of safety in terms of the probability of failure has been suggested by introducing the *reliability index* (RI) as the inverse of the *coefficient of variation* (COV) of the *limit-state function* (LSF) [7–10].

In the 1970s and later, the limit-state theory began to gradually replace the allowable or working stress design in structural codes of practice [11,12], leading to the development of a prescriptive approach called the *Load and Resistance Factor Design* (LRFD) in the United States [9,13–17], and the *semi-probabilistic* or *partial safety factors* (PSFs) design in Europe.

By introducing the semi-probabilistic design concept in the codes of practice, a problem has been aroused about how to reach the *target reliability index* (TRI) with the same PSFs of action for different resistance models and materials. The problem has been solved repeatedly by the calibration of PSFs of actions using a general "linear LSF" [10,18–24]. The calibration has provided an ease-of-use solution for the application of the partial safety method on the linear structural system, but those coded PSFs are not directly suitable for nonlinear structural systems.

One of the earliest thoughts on the influence of nonlinearity on structural safety has been considered by Basler [25], Figure 1. Balser has considered the example of a bending beam with an additional normal force. He distinguished between two cases: in the first case, the beam is loaded additionally by a tensile force $P$, and in the second case by a compressive force $P$. Considering the bending moment $M_m$ at the mid-span of the beam as the decisive effect of action, a concave curve is produced representing $P$ as a function of $M_m$ for the first case and a convex curve for the second case. Using what he called the "*container analogy*", Balser tried to introduce the concept of RI "*safety zone*" and its relation to the *global safety factor* (GSF). He has emphasized that only in cases where there is no proportionality between action and effect of action, it is necessary to distinguish whether the GSF is applied to the action or the effect of action.

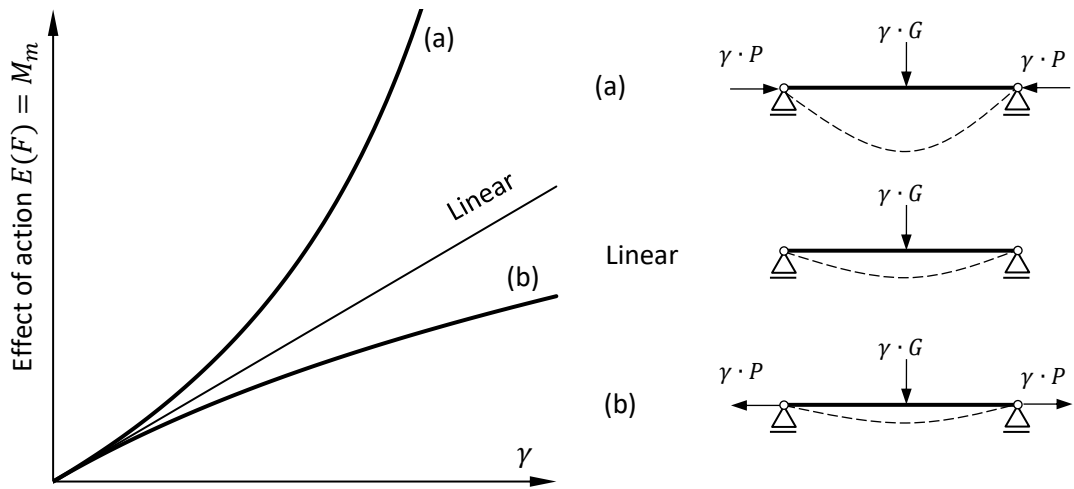

*Figure 1*     Relationship between the action increasing factor $\gamma$ and the effect of actions (the bending moment at the mid-span of the beam considering the geometric nonlinearity).

The example of Balser has been repeatedly considered by several researchers [26–28] as an application for the sub-clauses 6.3.2 (4) and 6.4.3.2 (4) of EN 1990:2002-10 [29]. The example provides significant intuition into the effects of nonlinearity on structural safety, but it also requires simultaneous increases in the normal and transversal actions.

The classification of the nonlinear behaviour as an *over-proportional* (OP) or *under-proportional* (UP) has been introduced in [30], sub-clauses 6.3.2 (4) and 6.4.3.2 (4) of EN 1990:2002-10 [29] as well as in the clause 8.3.2.1 of prEN 1990-2020-09 [31]. The purpose of classification is to provide simplified safe-sided rules to apply the PSF of action $\gamma_F$ to the action in the case of OP and to the effect in the case of UP.

According to [32], the mathematical definition of OP and UP can be described as follows:

(a) For OP behaviour:
$$E(\gamma_F \cdot F_k) > \gamma_F \cdot E(F_k) \tag{1}$$

(b) For UP behaviour:
$$E(\gamma_F \cdot F_k) < \gamma_F \cdot E(F_k) \tag{2}$$

Accordingly, the linear behaviour must satisfy the following condition:

$$E(\gamma_F \cdot F_k) = \gamma_F \cdot E(F_k) \tag{3}$$

where $E$ is the effect of the action and $F_k$ the characteristic value of the action. It is important to note that eq. (3) represents a condition for linear homogeneity.

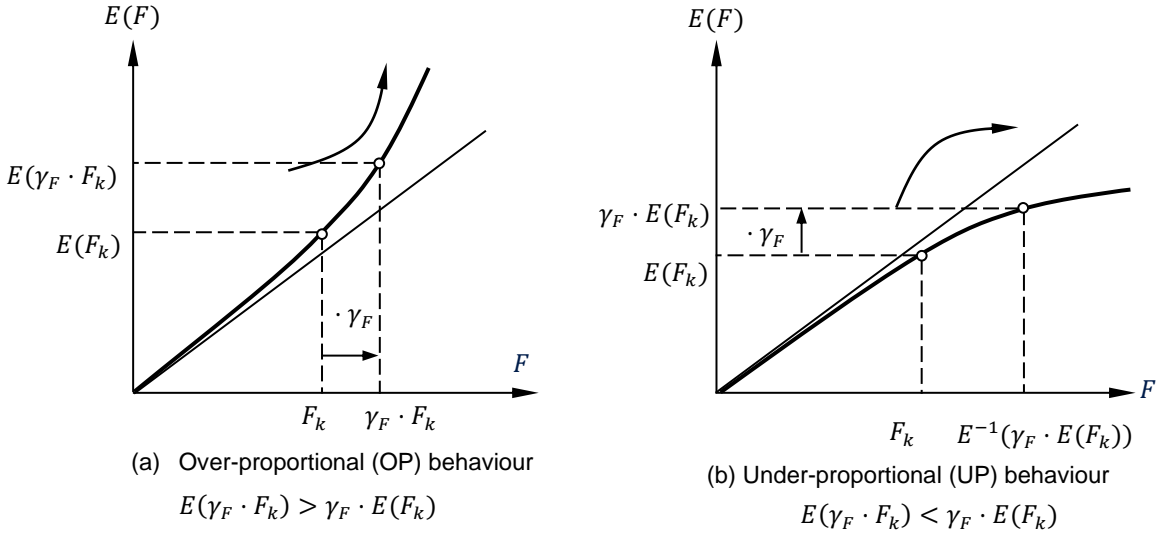

Figure 2     Application of the partial factor $\gamma_F$ in the case of non-linear analysis (single action). (a) The non-linear action effect increases more than the linear action effect proportional to the action. (b) The non-linear action effect increases less than the linear action effect proportional to the action [32]

Since the 1960s, there have been numerous inconsistencies in safety formats for nonlinear analyses [33]. Even though reliability and numerical simulation methods have made significant progress, the true character of the application of safety factors in non-linear structural systems has remained obscure in codes of practice, and their application is restricted to specific materials or structural elements, mainly reinforced concrete structures.

In the early studies of the reliability of nonlinear structural systems, over-safe structures have been obtained by applying coded PSFs [34]. This has led later to the recommendation for using the GSF format to achieve the desired probability of failure [35–38]. It has been suggested that the global resistance factor method could be applied to reinforced concrete members by approximating the resistance based on a two-parameter lognormal distribution [39,40]. The COV has been then estimated based on the mean value and characteristic value of the resistance. Using a constant sensitivity factor for resistance, the method of estimating the coefficient of variation (ECOV) [41,42] has been suggested to calculate the global safety factor of resistance. There have been many other researchers who have studied the global resistance format of reinforced concrete members, especially the M-N interaction behaviour [38,43–56]. However, by assuming a constant sensitivity of the resistance, the ECOV method remains restricted to specific applications of reinforced concrete structures and can't be used for general nonlinear applications. A study about the buckling of masonry walls in [57] showed that uncertainty in the elastic stiffness parameters (e.g. elastic modulus) affects both structural and resistance models. Using the global safety factor of resistance here reduces not only the material strengths but also the stiffnesses. In structures with large deformations, such as spatial, membrane, and cable structures [58,59], not only is nonlinearity more complicated but also the evaluation of safety becomes unclear. The same applies to geotechnical problems [60].

According to the preceding literature study, the impact of the nonlinearity of the structural system on safety has not been characterized for general applications. Each problem, structure, and material has been addressed independently. The current state of knowledge in this field still adheres to the recommendations of sub-clauses 6.3.2 and 6.4.3.2 (4) of EN 1990:2002-10 [29] with regards to the classification of nonlinearities as OP or UP [26,61–64].

This unclear treatment of nonlinearity may lead in many cases to unsafe designs, and in other cases to excessive unnecessary over-safety. With more actions applied to the structure, the problem becomes more complicated, and the OP-UP approach is no longer valid [32]. To treat the problem, a novel and general methodology named "*the theory of homogeneity*" has been proposed to characterize the nonlinearity of structural systems and to provide a mathematically proven basis for applying PSFs to nonlinear structural systems. The theory attempts to establish the key relationship between the PSF concept and the reliability theory of nonlinear structural systems by using the features of homogeneity.

## 2 Characterization of nonlinear structural systems

For intuitive understanding, the function $E = E(\mathbf{F})$ of the structural model is going to be considered in the next formulations, however, the same concept can be generalized and extended to the resistance model $R = R(\mathbf{M})$ or more generally to the LSF $g = g(\mathbf{X})$.

### 2.1 Homogeneity in nonlinear systems

In linear algebra, the function $E = E(\mathbf{F})$ is linear if it satisfies the following two conditions [65]: (1) Additivity or superposition principle, (2) Homogeneity of degree 1, or linear homogeneity, can be expressed as follows:

$$E(\gamma \mathbf{F}) = \gamma E(\mathbf{F}) \qquad (4)$$

The linear structural system has significant features concerning the PSFs, as it preserves the COV and the PSFs, i.e. $V_E = V_F$, $\gamma_E = \gamma_F$. These preservations cannot be achieved in a nonlinear structural system, but significant equivalent simplifications can be obtained if the nonlinear function $E = E(\mathbf{F})$ is *homogeneous*.

The function $E = E(\mathbf{F})$ is homogeneous of degree $n_E$, if satisfies the following identity ([66], p. 287):

$$E(\gamma \mathbf{F}) = \gamma^{n_E} E(\mathbf{F}) \qquad (5)$$

where $n_E$ is the *degree of homogeneity* (DH). Given $n_E = 1$, the function returns to the *linear homogeneity* form of eq. (4). If $n_E > 1$ the behaviour is OP or equivalent to eq. (1), and if $0 < n_E < 1$ the behaviour is UP or equivalent to eq. (2). This treatment suggests that $n_E$ can be used as a measure of nonlinearity if the nonlinear function $E(\mathbf{F})$ is homogenous. Although the function $E(\mathbf{F})$ is not homogeneous in general, it can be homogenized at the design point. A nonlinear structural system can be homogenized for this purpose by the concept introduced in section 2.2.

## 2.2 Homogenization of the nonlinear structural system

The homogenization of a nonlinear function $E(\mathbf{F})$ with $N_F$ actions can be approached similarly to linearization using the first terms of Taylor's series expansion after mapping the nonlinear function to a log-space. Considering that the action and resistance in many structural engineering applications have always been defined as positive values, it is possible to represent the nonlinear function $E = E(\mathbf{F})$ in log-space by using the following log-mapping ( $e = \ln E$ and $f_i = \ln F_i$).

Based on Taylor's series expansion around the mapped design point $\mathbf{f} = \mathbf{f}_d = [f_{d_i}] = [\ln F_{d_i}]$ of the new log-mapped function $e = e(\mathbf{f})$, the following equation is obtained:

$$e = e_d + \sum_{i=1}^{N_F} \frac{\partial e(\mathbf{f}_d)}{\partial f_i}(f_i - f_{d_i}) + h \tag{6}$$

where $e_d = \ln E_d$ and $h$ is the *reminder of homogenization approximation:*

The *Leibniz chain rule* can be used at the design point to get the following equation:

$$\frac{\partial e(\mathbf{f}_d)}{\partial f_i} = \frac{F_{d_i}}{E_d} \cdot \frac{\partial E(\mathbf{F}_d)}{\partial F_i} \tag{7}$$

By substituting eq. (7) into eq. (6), ignoring the remainder term, and returning to the original space of $E$ and $F_i$, it yields the following result:

$$\ln \frac{E}{E_d} = \sum_{i=1}^{N_F} n_{F_i} \ln \frac{F_i}{F_{d_i}} \tag{8}$$

where the term $n_{F_i}$ denotes the *partial degree of homogeneity* (PDH) of the function $E(\mathbf{F})$ for the action $F_i$ at the design point $\mathbf{F}_d$:

$$n_{F_i} = \frac{F_{d_i}}{E_d} \cdot \frac{\partial E(\mathbf{F}_d)}{\partial F_i} \tag{9}$$

Let us suppose there is a point $m$ between design action $F_{d_i}$ and action $F_i$ at which the function $E = E(\mathbf{F})$ is evaluated. If there is no correlation between the actions, it is possible to write the remainder of Taylor's series expansion eq. (6), in *Lagrange's form*, as follows:

$$h = \frac{1}{2!}\sum_{i=1}^{N_F}\sum_{j=1}^{N_F} H_{ij} \left(\ln \frac{F_i}{F_{d_i}}\right)\left(\ln \frac{F_j}{F_{d_j}}\right) = \frac{1}{2}\mathbf{f}^T \mathbf{H} \mathbf{f} \tag{10}$$

The matrix $H_{ij}$ is given as follows:

$$\mathbf{H} = [H_{ij}], \quad H_{ij} = n_{F_i F_j} + n_{F_i}(\delta_{ij} - n_{F_j}) \tag{11}$$

where $\delta_{ij}$ is the *Kronecker delta*. The terms $n_{F_i F_j}$ and $n_{F_i}$ are evaluated at $F_{m_i}$

$$n_{F_i F_j} = \frac{\partial^2 E}{\partial F_i \partial F_j} \frac{F_i F_j}{E}; \quad with \quad F_{m_i} = F_{d_i} \left(\frac{F}{F_{d_i}}\right)^c ; \quad c \in (0,1) \tag{12}$$

Since $\mathbf{H} \to 0$ implies that there is no reminder, it suggests that the approximation can be evaluated based on $\mathbf{H}$ matrix.

As a result of homogenisation, the effect $E = E(\mathbf{F})$ can be expressed in the following form:

$$E = E_d \prod_{i=1}^{N_F} \left(\frac{F_i}{F_{d_i}}\right)^{n_{F_i}} \tag{13}$$

Consequently, the resulting function in eq. (13) is now homogenous and meets the homogeneity definition in eq. (5).

*Euler's homogeneous function theorem* offers a compelling description of homogeneous functions [66]. The DH at the design point can be determined according to Euler's homogeneous function theorem as follows:

$$n_E \, E(\mathbf{F}_d) = \sum_{i=1}^{N_F} F_{d_i} \frac{\partial E(\mathbf{F}_d)}{\partial F_i} \tag{14}$$

As a result of the above equation, the DH can be calculated as follows:

$$n_E = \sum_{i=1}^{N_F} \frac{F_{d_i}}{E_d} \frac{\partial E(\mathbf{F}_d)}{\partial F_i} \tag{15}$$

Considering eq. (9), the following identity can be written between the DH and the PDHs for each action:

$$n_E = \sum_{i=1}^{N_F} n_{F_i} \tag{16}$$

Eqs. (5) and (8) can be extended to the resistance model with $N_M$ parameters. However, the subscript $E$ must be replaced with the subscript $R$ and the subscript $F$ must be replaced with the subscript $M$ as follows:

$$R(\gamma \mathbf{M}) = \gamma^{n_R} R(\mathbf{M}) \tag{17}$$

$$\ln \frac{R}{R_d} = \sum_{i=1}^{N_M} n_{M_i} \ln \frac{M_i}{M_{d_i}} \tag{18}$$

## 2.3 Partial safety factors (PSFs)

The relationship between the PSFs of actions and the PSF of the effect of actions is a direct result of the homogeneity property of the nonlinear function $E = E(\mathbf{F})$. Substituting the characteristic values of the actions $\mathbf{F} = \mathbf{F}_k$ in eq. (8) gives:

$$\ln \frac{E_k}{E_d} = \sum_{i=1}^{N_F} n_{F_i} \cdot \ln \frac{F_{k_i}}{F_{d_i}} \tag{19}$$

The design value of each unfavourable action $F_{d_i}$ can be written in terms of the characteristic value $F_{k_i}$ and the PSF of action $\gamma_{F_i}$, as follows: $F_{d_i} = \gamma_{F_i} F_{k_i}$. The same applies to the effect of actions $E_d = \gamma_E E_k$. By substituting $F_{d_i}$ and $E_d$ in eq. (19) gives:

$$\ln \gamma_E = \sum_{i=1}^{N_F} n_{F_i} \ln \gamma_{F_i} \tag{20}$$

Eq. (20) can also be written in the following form:

$$\gamma_E = \prod_{i=1}^{N_F} \gamma_{F_i}^{n_{F_i}} \tag{21}$$

Eq. (21) represents the relationship between the PSFs of actions $\gamma_{F_i}$ and the PSF of the effect of actions $\gamma_E$ in terms of the PDHs of the actions $n_{F_i}$.

The same also applies to determining the resistance PSF $\gamma_R$ in terms of the PSFs of the basic variables of the resistance model.

$$\gamma_R = \prod_{i=1}^{N_M} \gamma_{M_i}^{n_{M_i}} \tag{22}$$

## 2.4 Degree of homogeneity (DH) as a nonlinearity measure

It is convenient to consider a structural system with only one action to gain an intuitive sense of the DH. The DH, according to eq. (9), is defined as the ratio of the relative change in the effect of the action to the relative change in the action at the design point:

$$n = \frac{F_d}{E_d} \frac{dE(F_d)}{dF} \approx \frac{\Delta E/E_d}{\Delta F/F_d} = \frac{Relative\ change\ in\ the\ effect}{Relative\ change\ in\ the\ action} \tag{23}$$

Based on the value of $n$, the following cases can be distinguished considering both the effect and the action are positive, and the PSF of action $\gamma_F > 1$:

Case 1: $n > 0$ this happens only if the function $E = E(F)$ is monotonically increasing or the action and the effect are positively correlated at the design point. It means $E_d > E_k$. In terms of safety, there is an *unfavourable effect* associated with this action, and it is expected to result in $\gamma_E > 1$. As $n$ increases the influence of the action on structural safety increases (Figure 3). For this case, the nonlinear behaviour can be classified in terms of OP and UP, as follows:

(a) for $n > 1$, the behaviour is OP, which also can be called *over-linearly homogenous*. This implies that $\gamma_E > \gamma_F$.
(b) for $0 < n < 1$, the behaviour is UP, which also can be called *under-linearly homogenous*. This implies that $\gamma_E < \gamma_F$.
(c) for $n = 1$, the behaviour is equivalent to the *linear homogeneity*, but not necessary to be linear in general. This implies that $\gamma_E = \gamma_F$.

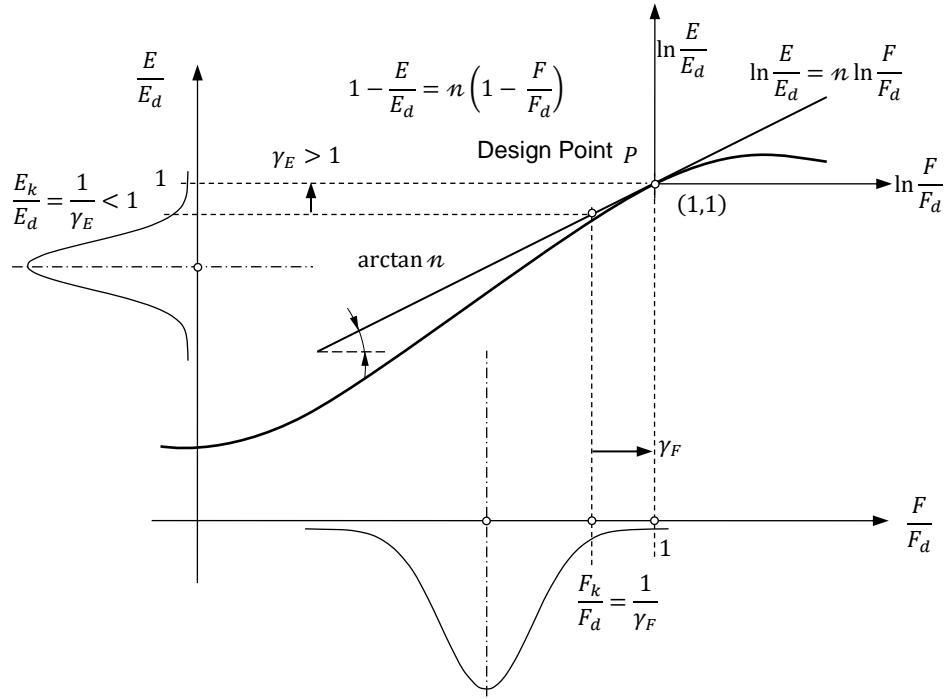

*Figure 3    Interpretation of the DH considering an unfavourable action.*

Case 2: $n < 0$ this happens only if the function $E = E(F)$ is monotonically decreasing or the action and the effect are negatively correlated at the design point. It means $E_d < E_k$. In terms of safety, there is a *favourable effect* associated with this action, and it is expected to result in $\gamma_E < 1$. This case can also be treated similarly to case 1 by changing the direction of action.

Case 3: $n = 0$ it happens if the function $E = E(F)$ remains constant in the vicinity of the design point. It means $E_d = E_k$. The action does not influence structural safety, and it is expected to have $\gamma_E = 1$.

In general, the DH is a value associated with the design point and may change from one point to another (Figure 4). However, the nonlinear system is homogeneous if the DH at all points remains constant.

Since the actions and material parameters are provided in the codes of practice mainly as characteristic values, this suggests calculating the DH based on the response of the nonlinear structural system to the characteristic and design values. Thus, the DH of the action $F$ at the design point can be approximated as follows:

$$n_F = \frac{1}{\ln \gamma_F} \ln \frac{E(\gamma_F F_k)}{E(F_k)} \tag{24}$$

In the case of multiple actions, the PDH can be approximated as follows:

$$n_{F_i} = \frac{1}{\ln \gamma_{F_i}} \ln \frac{E(\gamma_{F_1} F_{k_1}, \gamma_{F_2} F_{k_2}, \ldots, \gamma_{F_n} F_{k_n})}{E(F_{d_1}, F_{d_2}, \ldots, F_{k_i}, \ldots, F_{d_n})} \tag{25}$$

where $\gamma_{F_i}, F_{k_i}, F_{d_i}$ are the PSF, the characteristic value, and the design value of the action $F_i$, respectively.

The DH of the effect can be obtained directly by increasing all actions with the same factor $\gamma$ (like the example of Figure 1) as follows:

$$n_E = \frac{1}{\ln \gamma} \ln \frac{E(\gamma \mathbf{F})}{E(\mathbf{F})} \tag{26}$$

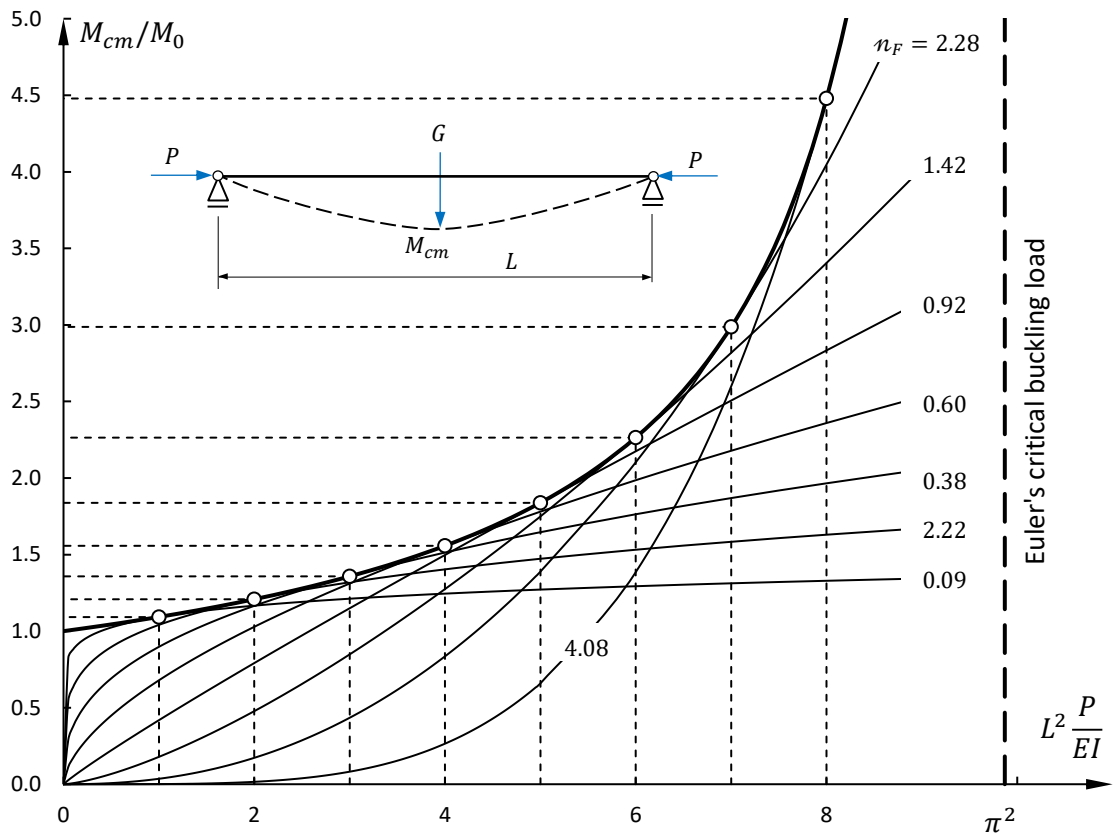

Figure 4    The variation of the DH at different design points of the non-linear Basler's example. The thick continuous curve illustrates the relationship between the action and the effect. The force $G$ is applied first and has negligible uncertainty while $P$ is a variable force. The thin curves correspond to the approximated homogeneous functions at the design points.

It is not sufficient to determine the PSF of the effect $\gamma_E$ based on the DH alone, but all PDHs for all actions must be calculated. A clear illustration of this can be found in eq. (21). However, for actions with different PSFs $\gamma_{F_1}, \gamma_{F_2}, \ldots, \gamma_{F_n}$, an *equivalent PSF* $\gamma_{eq}$ can be introduced, so that factoring all actions with $\gamma_{eq}$ gives the same effect as factoring each action with its PSF. If this factor exists, it must satisfy the following condition:

$$\gamma_E = \gamma_{eq}^{n_E} \tag{27}$$

By comparing eq. (27) with eq. (21), gives:

$$\gamma_{eq}^{n_E} = \prod_{i=1}^{N_F} \gamma_{F_i}^{n_{F_i}} \tag{28}$$

The equivalent PSF can be determined as follows:

$$\gamma_{eq} = \prod_{i=1}^{N_F} \gamma_{F_i}^{\nu_{F_i}} \tag{29}$$

where $\nu_{F_i}$ is the *relative partial degree of homogeneity* (RPDH) for the action $i$, and given by:

$$\nu_{F_i} = \frac{n_{F_i}}{n_E}; \quad \sum_{i=1}^{N_F} \nu_{F_i} = 1 \tag{30}$$

In the case of one action $F$, $n_E = n_F$ and $\nu_F = 1$, but for multiple actions with different PSFs, the equivalent PSF $\gamma_{eq}$ can only be evaluated if the RPDH $\nu_{F_i}$ are determined. It is not possible to create a useful relationship between the PSF of effect and PSFs of actions based on the OP-UP classification alone. Therefore, using the OP-UP classification for nonlinear systems with multiple actions may lead to an incorrect safety estimation.

## 3 Homogeneity analysis

According to sections 2.2 and 2.4, *homogeneity analysis* of the nonlinear structural system must be conducted to determine the DH, PDHs, and RPDHs at one or multiple design points. Based on these results, the relationship between the PSFs of actions and the PSF of effect can be derived directly from eq. (21) based on PDHs or from eqs. (27) and (28) based on DH and RPDHs. This can also be applied to resistance models to establish the relationship between the PSFs of resistance parameters and the PSF of resistance using eq. (22). The homogeneity analysis can also provide valuable insights about the points at which the DH takes specific transition or characteristic values like 0 or 1. Consequently, it facilitates the choice of design cases and the application of PSFs becomes straightforward even in complex nonlinear structural systems.

The homogeneity analysis can be efficiently integrated with advanced computational methods available in structural mechanics. A separate paper can be devoted to the background of the implementation of homogeneity analysis in computer codes. However, an introduction to the basis of homogeneity analysis can be provided by examining some simple nonlinear structural systems.

## 3.1 Flexural buckling of a column

An elastic column with simple supports on both sides has been considered in (Figure 5). The bending stiffness is $EI$ and length is $l$. The column is subjected to an eccentric compression load $P$ with a constant eccentricity $e$ at the top and bottom. Let's consider $M_m$ as the bending moment at the middle of the column $M_m$ and $P_E$ as the *Euler buckling load* $P_E = \frac{\pi^2}{l^2} EI$. By representing the action and the effect in relative form $\eta = M_m/P_E\, e$ ; $\xi = P/P_E$, respectively, the relationship between the effect $\eta$ and action $\xi$ can be described based on the nonlinear *second-order analysis* as follows:

$$\eta = \frac{\xi}{\cos\left(\frac{\pi}{2}\sqrt{\xi}\right)} \tag{31}$$

By application of eq. (9), the DH at action parameter $\xi$ is given as follows:

$$n = 1 + \frac{1}{2}\alpha \tan(\alpha) > 1; \text{with } \alpha = \frac{\pi}{2}\sqrt{\xi} \tag{32}$$

Since the action parameter $\xi$ is limited between $0 \leq \xi \leq 1$, the parameter $\alpha$ is also limited between $0 \leq \alpha \leq \pi/2$, therefore, the DH $n$ is bigger than 1 under any compression action.

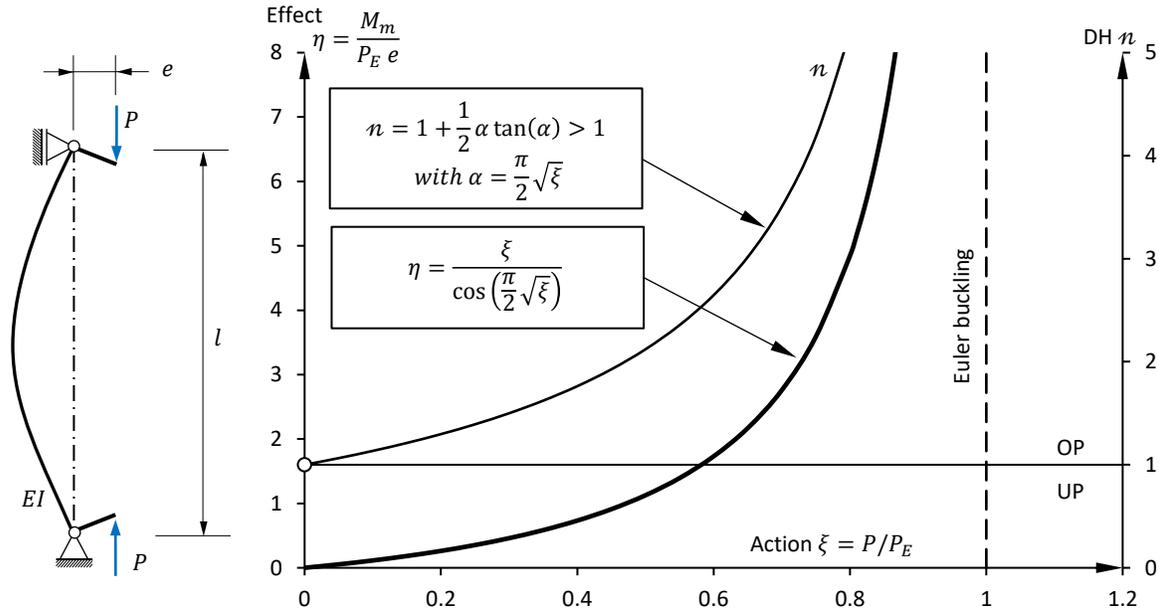

*Figure 5*    *The action-effect behaviour of the flexural buckling of a column under an eccentric compression action is based on nonlinear second-order analysis. The relationship between the action and the DH is shown on the same diagram and it is based on the homogeneity analysis.*

## 3.2 Bending under eccentric tension

The same example in section 3.1 has been considered once again but with eccentric tension load $P$ in (Figure 6). Let's consider $M_m$ as the bending moment at the middle of the column $M_m$ and $P_E$ the Euler buckling load $P_E = \frac{\pi^2}{l^2} EI$. By representing the action and the effect in relative form $\eta = M_m/P_E\, e$ ; $\xi = P/P_E$, respectively, the relationship between the effect $\eta$ and action $\xi$ can be described based on the nonlinear second-order analysis as follows:

$$\eta = \frac{\xi}{\cosh\left(\frac{\pi}{2}\sqrt{\xi}\right)} \tag{33}$$

By application of eq. (9), the DH at action parameter $\xi$ is given as follows:

$$n = 1 - \frac{1}{2}\alpha \tanh(\alpha)\, ; with\ \alpha = \frac{\pi}{2}\sqrt{\xi} \tag{34}$$

From eq. (34), it can be noted that the DH $n$ is less than 1 under any tension action. However, as tension increases, the bending moment increases up to a certain limit $\eta_{max} = 0.431$, after which the bending moment starts to decrease. The maximum effect can be determined by setting $n = 0$ in eq. (34). This corresponds to $\xi = 1.729$. In the ascending branch $0 < n < 1$ the action and effect are positively correlated (case 1), however in the descending branch $n < 0$ the action and effect become negatively correlated (case 2).

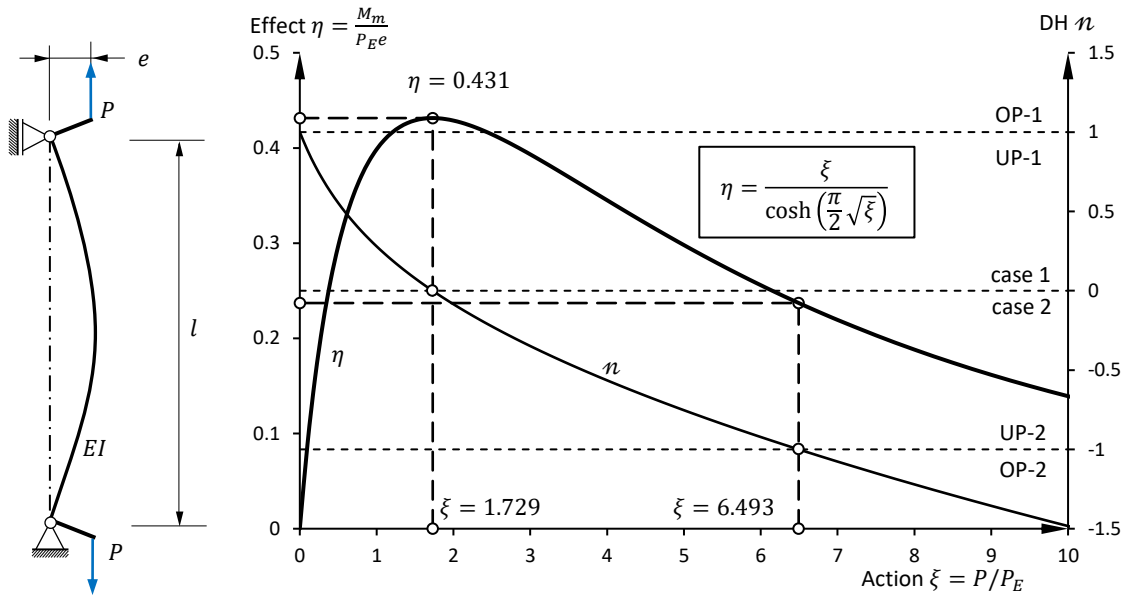

*Figure 6*     *The nonlinear action-effect behaviour of an element under an eccentric tension action is based on second-order analysis. The relationship between the action and the DH is shown on the same diagram and it is based on the homogeneity analysis.*

## 3.3 Cable element with lateral force

Let's consider a cable of axial stiffness $EA$ and an initial length of $2l$. The cable is fixed on two supports with a distance of $2l$ in between. Let's apply a lateral force $P$ in the middle and determine the tensile force $N$ in the cable. By representing the action and the effect in relative form $\eta = N/EA$ ; $\xi = P/EA$ , the relationship between the effect $\eta$ and action $\xi$ can be described based on the nonlinear large-deformation analysis as in (Figure 6-a).

Figure 6-b shows the action-DH relationship obtained from homogeneity analysis according to eq. (9). It is evident from the homogeneity analysis that the variation of DH is limited within the range $2/3 \leq n < 1$, which indicates relatively a *small nonlinearity*.

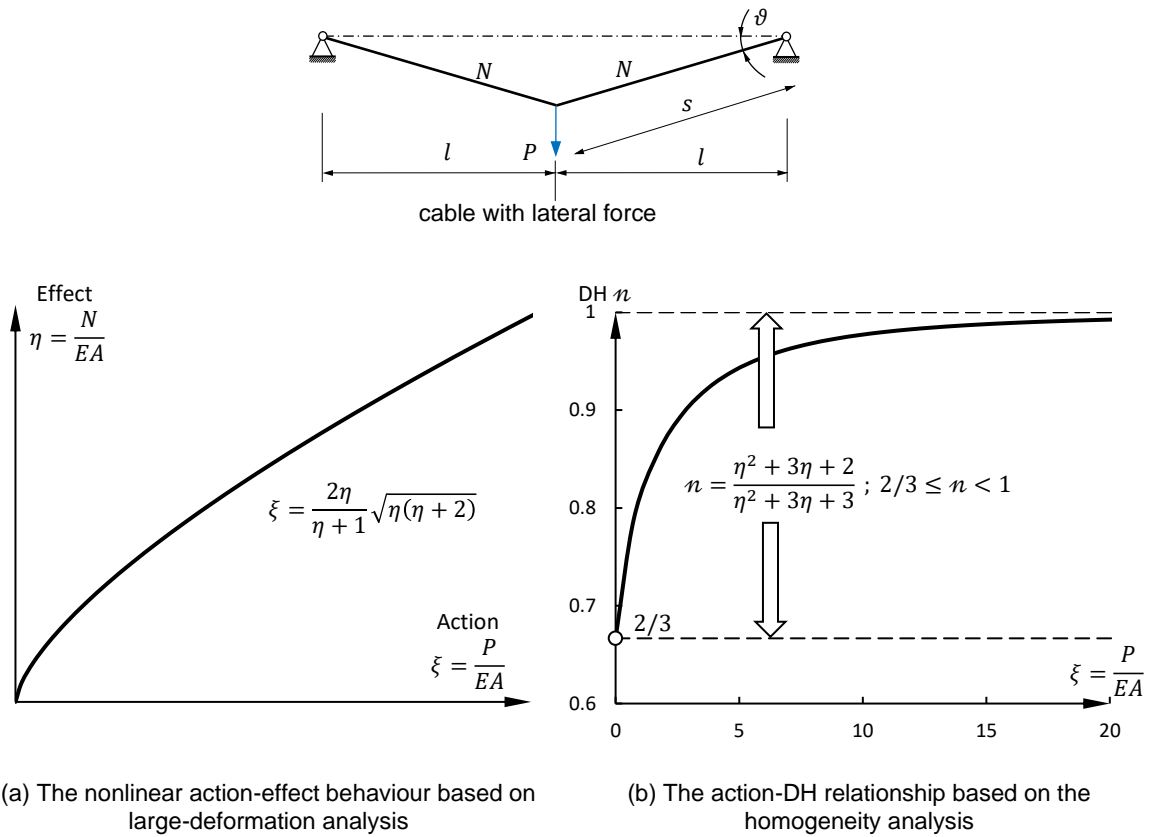

(a) The nonlinear action-effect behaviour based on large-deformation analysis

(b) The action-DH relationship based on the homogeneity analysis

*Figure 7*   *The results of large-deformation analysis and homogeneity analysis of a cable element with one transversal action.*

## 3.4 Beam with two actions

The beam with two actions has already been considered in Figure 1. In the current example, the transversal load $G$ is applied first, followed by the normal force $P$. Let $M_0 = Gl^2/4$ be the bending moment at the mid-span of the beam due to load $G$. With a load P applied, the mid-span bending moment of the beam becomes $M_m$. Considering the relative description of the action $\xi = P/P_E$ and effect $\eta = M_m/M_0$, where $P_E = \pi^2 EI/l^2$ is the Euler buckling load, the relationship between the effect $\eta$ and action $\xi$ can be described based on the nonlinear second-order analysis as in Figure 8. By performing a homogeneity analysis, the relationships between the action $\xi$ and the DH, PDHs, RPDHs can be obtained in Figure 9 and Table 1.

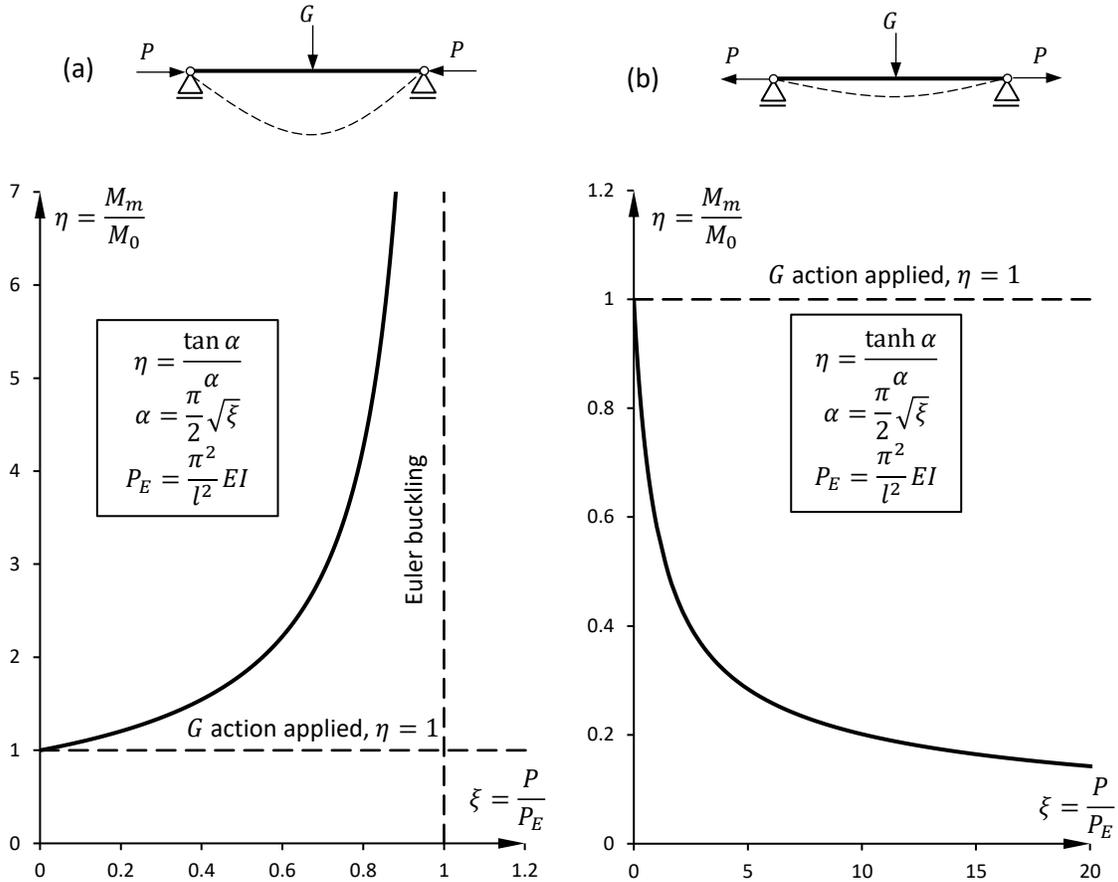

*Figure 8   The effect-action relationship for a beam with two actions ($G$ transversal and $P$ normal) is based on nonlinear second-order analysis. The effect represents the bending moment at the mid-span of the beam. First, the action $G$ is applied and increased to reach a moment $M_0 = Gl^2/4$ at the mid-span of the beam. Next, the action $P$ is applied and increased to reach a moment $M_m$ at the mid-span of the beam. (a) $P$ is a compression force, (b) $P$ is a tension force.*

*Table 1   Determination of the homogeneity characteristic parameters for the example of a beam with two actions*

| Parameter | (a) $P$ is a compression force | (b) $P$ is a tension force |
| --- | --- | --- |
| The effect $\eta$ | $\eta = \dfrac{\tan \alpha}{\alpha}$; $\alpha = \dfrac{\pi}{2}\sqrt{\xi}$ <br> $0 < \xi < 1$; $0 < \alpha < \pi/2$ | $\eta = \dfrac{\tanh \alpha}{\alpha}$; $\alpha = \dfrac{\pi}{2}\sqrt{\xi}$ <br> $\xi > 0$; $\alpha > 0$ |
| PDH for the action $G$, eq. (9) | $n_G = 1$ | $n_G = 1$ |
| PDH for the action $P$, eq. (9) | $n_P = \dfrac{1}{2}\left(\dfrac{2\alpha}{\sin 2\alpha} - 1\right)$ <br> $n_P > 0$ | $n_P = \dfrac{1}{2}\left(\dfrac{2\alpha}{\sinh 2\alpha} - 1\right)$ <br> $0 < n_P < -1/2$ |
| DH for the effect, eq. (16) | $n_E = \dfrac{1}{2}\left(1 + \dfrac{2\alpha}{\sin 2\alpha}\right)$ <br> $n_E > 1$ | $n_E = \dfrac{1}{2}\left(1 + \dfrac{2\alpha}{\sinh 2\alpha}\right)$ <br> $1/2 < n_E < 1$ |
| RPDH for the action $G$, eq. (28) | $\nu_G = \dfrac{2 \sin 2\alpha}{2\alpha + \sin 2\alpha}$ | $\nu_G = \dfrac{2 \sinh 2\alpha}{2\alpha + \sinh 2\alpha}$ |
| RPDH for the action $P$, eq. (28) | $\nu_P = \dfrac{2\alpha - \sin 2\alpha}{2\alpha + \sin 2\alpha}$ | $\nu_P = \dfrac{2\alpha - \sinh 2\alpha}{2\alpha + \sinh 2\alpha}$ |
| PSF of effect $\gamma_E$, eqs. (21), (29) | $\gamma_E = \gamma_{eq}^{n_E} = \gamma_G^{n_G}\gamma_P^{n_P}$; $\gamma_{eq} = \gamma_G^{\nu_G}\gamma_P^{\nu_P}$ | |

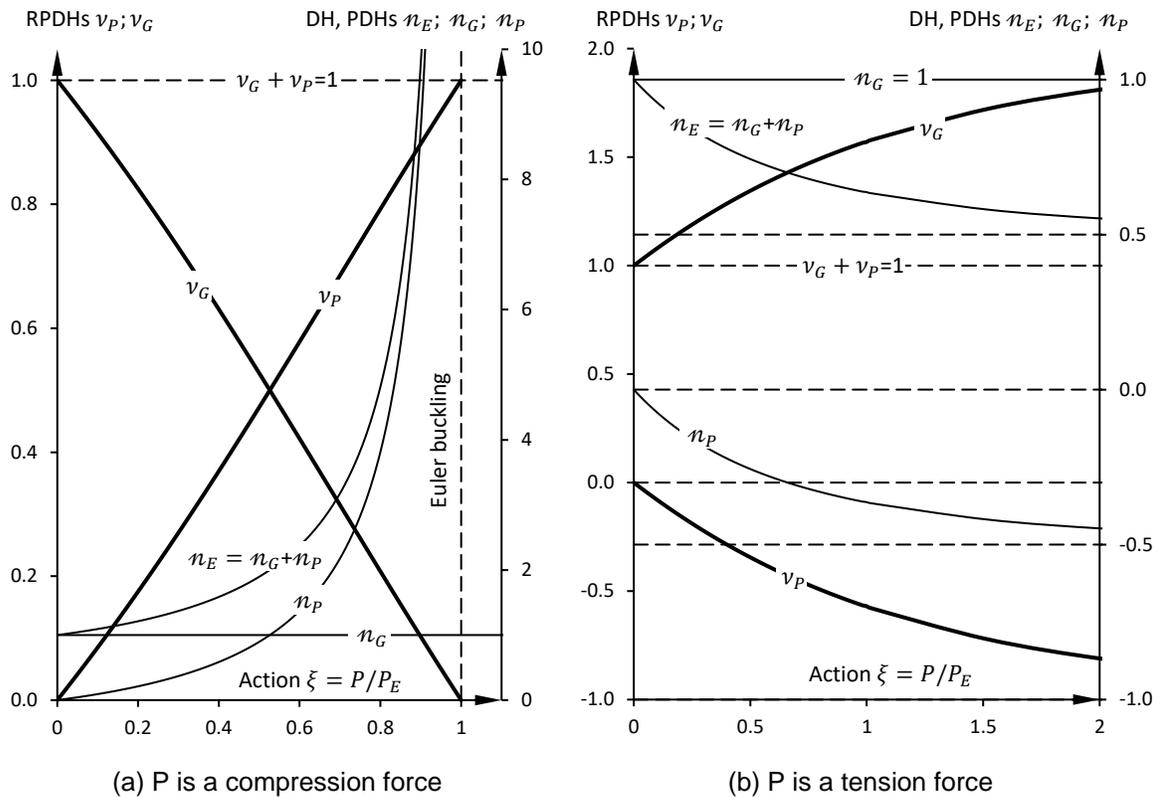

*Figure 9*  The relationship between the homogeneity characteristics (DH $n_E$, PDHs $n_G$, $n_P$, and RPDHs $\nu_P, \nu_G$) and the action variables $\xi$ based on homogeneity analysis for the beam example with two actions.

If P is a compression force, then $n_P > 0$ and it corresponds to unfavourable action. A PDH $n_P = 1$ can be obtained when $\xi = 0.526$. This means if $G$ remains constant with only $P$ being increased, this results in an UP behaviour if $\xi < 0.526$ and in an OP behaviour if $\xi > 0.526$. Since $n_E > 1$, it follows that increasing both forces $G$ and $P$ with the same factor produces an OP behaviour.

If P is a tension force, then $0 < n_P < -1/2$ and it corresponds to favourable action. Since $1/2 < n_E < 1$, it follows that increasing both forces $G$ and $P$ with the same factor produces an UP behaviour.

## 3.5 Masonry shear wall

Let's consider a masonry shear wall subjected to a vertical action $F_1$ (self-weight) and a horizontal load $F_2$ (wind). The compression stress $\sigma$ at the base of the wall can be calculated using the assumption of a *rectangular stress block* of the cracked cross-section. Given that the effect is the stress at the base of the wall, it can be expressed alternatively as $E = \sigma \cdot t \cdot l_w$. The effect-action relationships are demonstrated in Figure 10 for two cases, the first considers the action $F_1$ constant and the second consider the action $F_2$ constant.

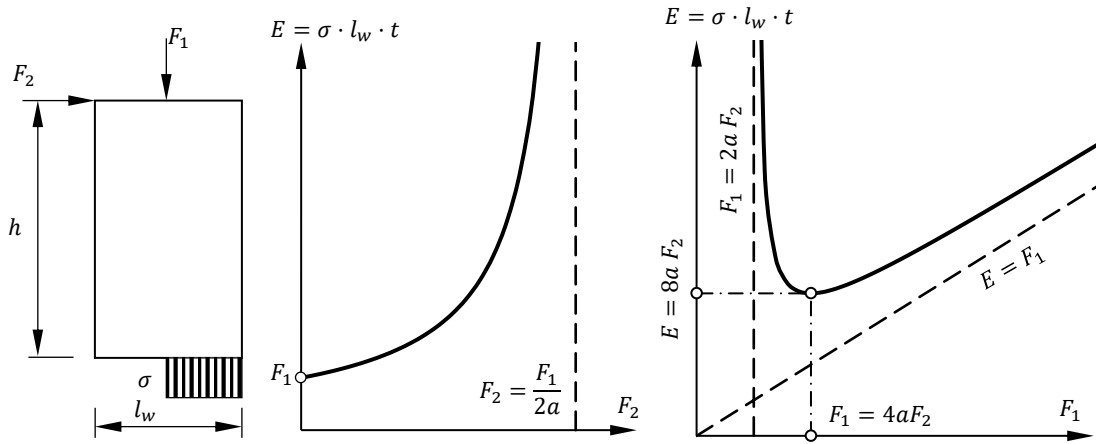

Figure 10  The effect-action relationship for a masonry shear wall with two actions ($F_1$ vertical and $F_2$ horizontal) based on stress block assumption.

By performing a homogeneity analysis, the relationships between the actions and the DH, PDHs, and RPDHs, respectively, can be obtained in Figure 11 and Table 2. It is evident from the homogeneity analysis of the shear wall in Table 2 that DH $n_E = 1$ is an example of a linear homogeneity. However, it does not imply that the system is linear. It means that an increase in both horizontal and vertical action in a masonry shear wall would also increase the effect (the stress at the base) by the same factor.

Table 2  Determination of the homogeneity characteristic parameters for the example of masonry shear wall

| | |
|---|---|
| The effect $\eta$ | $E(F_1, F_2) = \dfrac{F_1^2}{F_1 - 2aF_2}$ |
| PDH for the action $F_1$, eq. (9) | $n_{F_1} = \dfrac{F_1 - 4aF_2}{F_1 - 2aF_2}$ |
| PDH for the action $F_2$, eq. (9) | $n_{F_2} = \dfrac{2aF_2}{F_1 - 2aF_2}$ |
| DH for the effect $E$, eq. (16) | $n_E = n_{F_1} + n_{F_2} = 1$ |
| RPDH for the action $F_1$, eq. (28) | $\nu_{F_1} = n_{F_1} = \dfrac{F_1 - 4aF_2}{F_1 - 2aF_2}$ |
| RPDH for the action $F_2$, eq. (28) | $\nu_{F_2} = n_{F_2} = \dfrac{2aF_2}{F_1 - 2aF_2}$ |
| PSF of effect $\gamma_E$, eqs. (21), (29) | $\gamma_E = \gamma_G^{n_G} \gamma_P^{n_P} = \gamma_{eq} = \gamma_G^{\nu_G} \gamma_P^{\nu_P}$ |

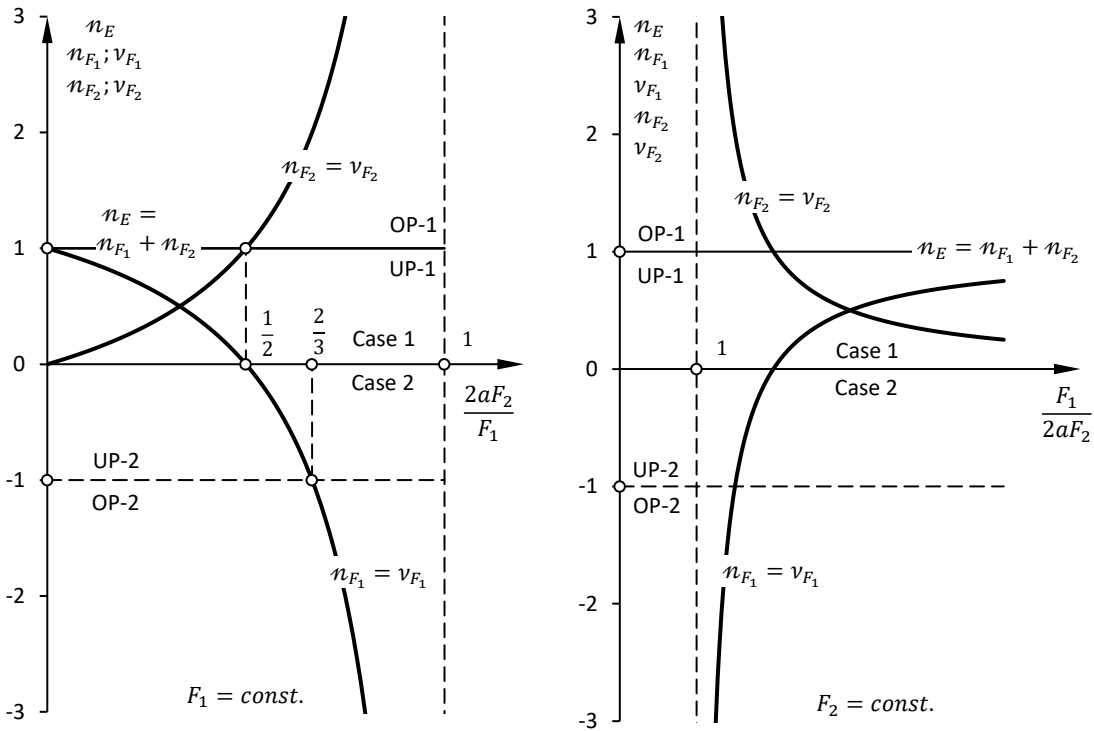

Figure 11  The relationship between the DH $n_E$, PDHs $n_G$; $n_P$, and RPDHs $\nu_P$; $\nu_G$ and the action variables for the shear wall example with two actions.

# 4 Nonlinearity and safety

## 4.1 RI of nonlinear LSF

Safety checks are performed at critical locations using predefined LSFs of a critical failure mode. The LSF may also be expressed in the following equivalent form at the design point using the *design value format*:

$$g(\mathbf{F}, \mathbf{M}, \theta_E, \theta_R) = \frac{R(\mathbf{M}, \theta_R)/R_d}{E(\mathbf{F}, \theta_E)/E_d} \tag{35}$$

To account for the model uncertainty, the theoretical models of resistance and effect can be updated by a *bias correction model* [67–71] as follows: $E(\mathbf{F}, \theta_E) = b_E\, E_t(\mathbf{F})\, \theta_E$ and $R(\mathbf{M}, \theta_R) = b_R\, R_t(\mathbf{M})\, \theta_R$, where $E_t(\mathbf{F})$ is the theoretically calculated effect of action, $b_E$ is the bias of the structural model, $\theta_E$ is the *error parameter* of the structural model, $R_t(\mathbf{M})$ is the theoretically calculated resistance, $b_R$ is the bias of the resistance model, $\theta_R$ is the error parameter of the resistance model.

Analogue to eqs. (8) and (18), the logarithm of the LSF in eq. (35) can be described at the design point as follows:

$$\ln g = \ln\frac{\theta_R}{\theta_{R_d}} + \sum_{i=1}^{N_M} n_{M_i} \ln\frac{M_i}{M_{d_i}} - \ln\frac{\theta_E}{\theta_{E_d}} - \sum_{i=1}^{N_F} n_{F_i} \cdot \ln\frac{F_i}{F_{d_i}} \tag{36}$$

Eq. (36) is suitable for the two-model format when two independent models for the resistance and the structural analysis are used. If each of the variables $M_i, F_i, \theta_E, \theta_R$ are denoted by a generalized variable $X_i$ the above equation can be extended to any multiple models or single-model format to become as follows:

$$\ln g = -\sum_{i=1}^{N} n_i\bigl(\ln X_i - \ln X_{d_i}\bigr) \tag{37}$$

Where $N$ is the total number of the basic variables including the error parameters in the models.

The status of the basic variable $X_i$ is defined as unfavourable at the design point if the increase of this variable produces an unfavourable effect on the safety of the system and is defined as favourable if the increase of this variable produces a favourable effect on the safety of the system.

**Definition 1** *The design value of the unfavourable variable $X_i$ is given as $X_{d_i} = \gamma_i\, X_{k_i}$, ($X_{k_i}$ the upper or superior characteristic value) and the design value of the favourable variable $X_i$ is given as $X_{d_i} = X_{k_i}/\gamma_i$ ($X_{k_i}$ the lower or inferior characteristic value).*

By using Definition 1, now both favourable and unfavourable variables can be treated equally in the formulation of RI. If the variable $X_i$ is lognormally distributed, based on eq. (37), the RI can be generalised to $N$ number of basic variables as follows:

$$\beta = \frac{\sum_{i=1}^{N} n_i(\ln \gamma_i + k_i Q_i)}{\sqrt{\sum_{i=1}^{N}(n_i Q_i)^2}} \; ; \; Q_i = \sqrt{\ln(1 + V_i^2)}; \; k_i = \Phi^{-1}(p_i) \tag{38}$$

where $p_i$ is the percentile at which the characteristic value is determined, $V_i$ is the COV of $X_i$, where $X_i$ may represent an action parameter, material parameter, or model error parameter (Note that the PDH for the error parameter of the model is 1).

Eq. (38) represents a significant outcome of the homogenization theory. The RI is described in a closed form in terms of the statistical parameter $Q_i$ ($Q_i$ is approximately equal to the COV when $V_i < 0.2$) and the PDH of each variable.

It is convenient to rearrange eq. (38) of the RI in the following form:

$$\beta = \frac{\sum_{i=1}^{N} n_i Q_i \beta_i}{\sqrt{\sum_{i=1}^{N} (n_i Q_i)^2}} \quad (39)$$

where $\beta_i$ can be defined as the *partial reliability index* (PRI) of the variable $i$. It is given as follows for the lognormal distribution:

$$\beta_i = k_i + \frac{1}{Q_i} \ln \gamma_i \quad (40)$$

Note that $\beta_i$ is an invariant parameter to nonlinearity, and it depends only on the statistical properties and the PSF of the basic variable $i$.

To understand the importance of the PRI, let's assume that the PDH $n_i$ of the basic variable $X_i$ is too large. As a result, the RI becomes highly sensitive to the basic variable $X_i$, but at the same time, its sensitivity to all other basic variables vanishes. This can be demonstrated by finding the limit of the RI in eq. (39) at an infinite PDH:

$$\lim_{n_i \to \pm\infty} \beta = \pm \frac{\ln \gamma_i + k_i Q_i}{Q_i} = \pm \beta_i \quad (41)$$

Considering the above treatment, the PRI $\beta_i$ of a basic variable $i$ may be defined as the RI with only the variable $i$ dominating the system.

If the variable $X_i$ has a non-lognormal distribution, the RI in eq. (39) takes the following generalized form:

$$\beta = \frac{\sum_{i=1}^{N} n_i \tau_i Q_i \beta_i}{\sqrt{\sum_{i=1}^{N} (n_i \tau_i Q_i)^2}} \quad (42)$$

where $\tau_i$ is the *degree of homogeneity of the non-lognormality* (DHN) at the design point and is given by:

$$\tau_i = Q_i \frac{g_i(X_{d_i})}{\phi\left(\Phi^{-1}\left(G_i(X_{d_i})\right)\right)} X_d \quad (43)$$

where $g_i(X_i)$ is the *probability density function* of the *non-lognormal distribution* and $G_i(X_i)$ is the *cumulative distribution function*. The DHN $\tau_i = 1$ for the lognormally distributed basic variables.

## 4.2  Upper and lower bounds of RI

The RI in eq.(42) can be expressed in the matrix form as follows:

$$\beta = \frac{1}{\|\mathbf{q}\|}\boldsymbol{\beta}^T\mathbf{q} \tag{44}$$

where the vector $\boldsymbol{\beta}$ is the vector of the PRI for all basic variables:

$$\boldsymbol{\beta} = [\beta_i] \tag{45}$$

The value of $\beta_i$ is *invariant to the nonlinearity*. It depends only on the distribution type and the statistical properties of the variable $i$.

The vector $\mathbf{q}$ is defined as follows:

$$\mathbf{q} = [q_i], q_i = n_i \tau_i Q_i \tag{46}$$

It represents the total contributions from all safety influencing factors, namely:

(1) the statistical parameter $Q_i$ or the COV;
(2) the statistical distribution of the basic variables, represented by the DHN $\tau_i$;
(3) the nonlinearity of the LSF, characterized by the PDH $n_i$.

The total sensitivity vector $\boldsymbol{\alpha}$ can be obtained by normalizing the vector $\mathbf{q}$, as follows:

$$\boldsymbol{\alpha} = \frac{1}{\|\mathbf{q}\|}\mathbf{q} \tag{47}$$

where $\|\boldsymbol{\alpha}\| = \sqrt{\boldsymbol{\alpha}^T\boldsymbol{\alpha}} = 1$. As a result of the above definition of the total sensitivity vector $\boldsymbol{\alpha}$, the RI of eq. (44) can be reduced into the following simple form:

$$\beta = \boldsymbol{\beta}^T\boldsymbol{\alpha} \tag{48}$$

An *upper bound* for the RI $\beta$ can be directly obtained from eq. (48), by applying the *Cauchy–Schwarz inequality* [72]:

$$\boldsymbol{\beta}^T\boldsymbol{\alpha} \leq \|\boldsymbol{\beta}\|\|\boldsymbol{\alpha}\| = \|\boldsymbol{\beta}\| \tag{49}$$

As a result, the RI has an upper bound of $\beta_{max}$:

$$\beta_{max} = \|\boldsymbol{\beta}\| = \sqrt{\boldsymbol{\beta}^T\boldsymbol{\beta}} = \sqrt{\sum_{i=1}^{N}\beta_i^2} \tag{50}$$

Optimal PSFs of variables $X_i$ are those which meet the upper bound of the RI. This can be reached if both vectors $\boldsymbol{\beta}$ and $\boldsymbol{\alpha}$ are parallel:

$$\frac{\beta_i}{\alpha_i} = const. \ for \ all \ i \tag{51}$$

Nevertheless, the semi-probabilistic method doesn't follow this optimisation principle as the PSFs are coded as constant values.

Eq. (50) means that the RI can reach only one unique peak. As we move away from the peak, the function $\beta$ decreases in all directions. The minimum values can then be reached at extreme sensitivities when $\alpha_i = 1$ $and$ $\alpha_{j \neq i} = 0$ which corresponds to $\beta = \beta_i$. The lower bound of the RI can be obtained as the minimum value in the vector $\boldsymbol{\beta}$, and can be expressed as follows:

$$\beta_{min} = \min(\beta_i) \tag{52}$$

By considering all possible nonlinearities, the RI has an upper and lower bound:

$$\sqrt{\boldsymbol{\beta}^T \boldsymbol{\beta}} \leq \beta \leq \min(\boldsymbol{\beta}) \tag{53}$$

This leads to the following *Bakeer's theorem 1 of the upper and lower bounds of the RI:*

> **Bakeer's theorem 1   The upper and lower bounds of the RI**
>
> If specific PSFs $\gamma_i \geq 1$ are applied to the basic variables $X_i$ of a nonlinear structural system according to Definition 1, the RI $\beta$ remains always bounded between:
>
> – an upper bound equal to $\sqrt{\boldsymbol{\beta}^T \boldsymbol{\beta}}$ and
> – a lower bound equal to $min(\boldsymbol{\beta})$

## 4.3   The influence of nonlinearity on RI

To understand the upper and lower bounds of the RI, let's consider a structural system in the two-model format with one material parameter $M$ in the resistance model and one action $F$ in the structural model. For given PSFs, $\gamma_F > 1$ and $\gamma_M > 1$, according to eq. (44), the RI can be written as follows:

$$\beta = \frac{q_M \cdot \beta_M + q_F \cdot \beta_F}{\sqrt{q_M^2 + q_F^2}} \tag{54}$$

Where $\beta_M$ and $\beta_F$ are the PRIs of material and action, respectively. Let's define the *relative sensitivity parameter* (RSP) $\xi$, which measures the ratio between the sensitivity of the action to the sensitivity of the material parameter:

$$\xi = \frac{q_F}{q_M} = \frac{n_F}{n_M} \frac{\tau_F}{\tau_M} \frac{Q_F}{Q_M} \tag{55}$$

In a special case when the resistance model is linear and has a lognormal distribution then $n_M = 1$; $\tau_M = 1$; $Q_M = Q_R$; $\beta_M = \beta_R$. Assuming $\tau_F n_F = n$; gives $\xi = n \frac{Q_F}{Q_R}$. The RI can be expressed as a function of $\xi$ as follows:

$$\beta(\xi) = \frac{\beta_R + \xi \cdot \beta_F}{\sqrt{1 + \xi^2}}; \quad \xi = n \frac{Q_F}{Q_R} \tag{56}$$

Figure 12 shows the relationship between the RI $\beta(\xi)$ and the RSP $\xi$. Note that $\beta_R$ is the RI at no sensitivity to action and it corresponds to $\xi = 0$. $\beta_F$ is the RI at no sensitivity to resistance and it corresponds to $\xi = \infty$. The upper bound of the RI can be reached at $\xi = \frac{\beta_F}{\beta_R}$; $\beta_m = \sqrt{\beta_F^2 + \beta_R^2}$, which occurs at the DH:

$$n = \frac{\beta_F/Q_F}{\beta_R/Q_R} \tag{57}$$

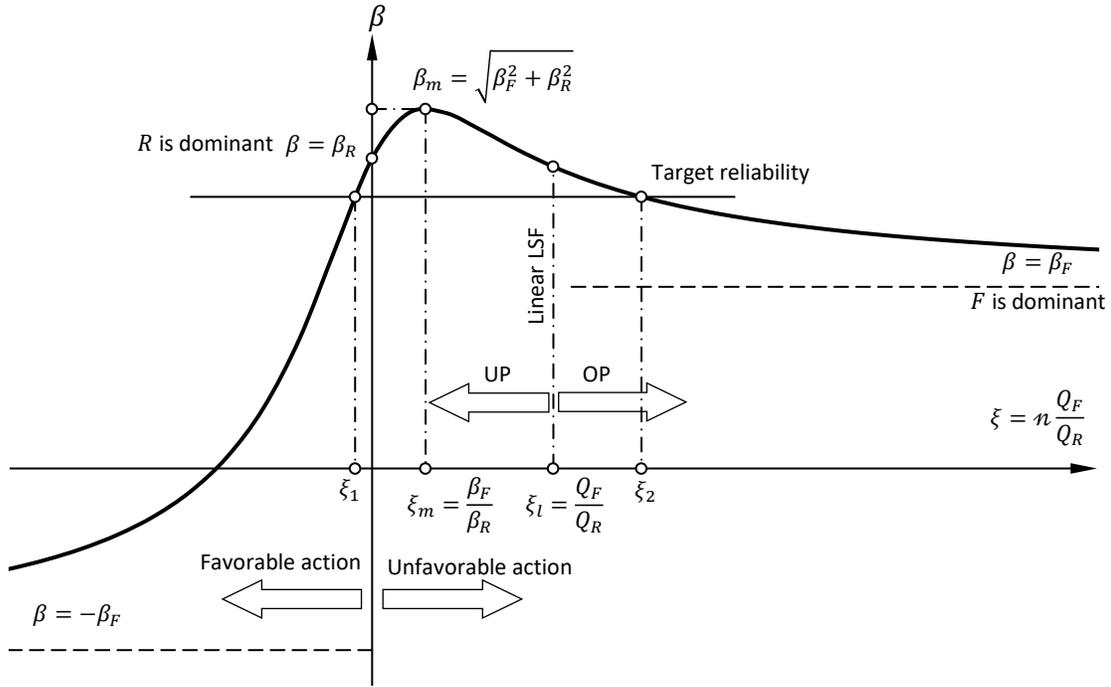

*Figure 12   The variation of the RI $\beta$ as a function of the RSP $\xi$ using a PSF of action $\gamma_F > 1$.*

The case $\xi \geq 0$ corresponds to unfavourable action. At this range, $\beta$ starts with $\beta_R$ at $\xi = 0$ and increases to $\beta_m = \sqrt{\beta_F^2 + \beta_R^2}$ at $\xi = \frac{\beta_F}{\beta_R}$ and then reduces to $\beta = \beta_F$ at $\xi = \infty$. Therefore, for $\xi \geq 0$, the lower bound of $\beta$ is $\min(\beta_R, \beta_F)$. This agrees with Bakeer's theorem 1, that in a nonlinear structural system with one action, the lower bound of RI can be calculated as $\beta_{min} = \min(\beta_R, \beta_F)$. This statement suggests that the PSF of action which is determined according to $min(\beta_R, \beta_F) = \beta_t$ should bring the target reliability level for any nonlinear system.

The case $\xi < 0$ corresponds to favourable action. The value of RI $\beta$ in this range decreases as the parameter $\xi$ decreases. This behaviour is directly explained by using a PSF of more than 1 for unfavourable action.

The linear case corresponds to $\xi = Q_F/Q_R$ (Figure 12). The values of $\xi$ at target reliability $\beta_t$ can be calculated as follows:

$$\xi_R = \frac{\beta_F \beta_R - \beta_t\sqrt{\beta_m^2 - \beta_t^2}}{\beta_t^2 - \beta_F^2}; \; \xi_F = \frac{\beta_F \beta_R + \beta_t\sqrt{\beta_m^2 - \beta_t^2}}{\beta_t^2 - \beta_F^2} \tag{58}$$

## 4.4 The influence of model uncertainty on RI

The Bakeer's theorem 1 of the upper and lower bound of the RI can also be demonstrated by considering a structural system with one action and one resistance parameter, but with model uncertainty. By using the RSPs $\xi_F = n\frac{Q_F}{Q_R}; \xi_\theta = \frac{Q_\theta}{Q_R}$, the RI $\beta = \beta(\xi_F, \xi_\theta)$ can be represented as a 3D surface or contour lines (Figure 13). As can be seen from the plotted surface, the RI reaches its upper bound $\beta_{max} = \sqrt{\beta_R^2 + \beta_F^2 + \beta_\theta^2}$ at one peak which corresponds to $\xi_{m_F} = \frac{\beta_F}{\beta_R}; \xi_{m_\theta} = \frac{\beta_\theta}{\beta_R}$. Moreover, the lower bound of the RI can be determined as a minimum of three values $\beta_{min} = \min(\beta_R, \beta_F, \beta_\theta)$. The model error parameter $\theta$ has been treated like the action and resistance parameters but considering that the PDH equals 1.

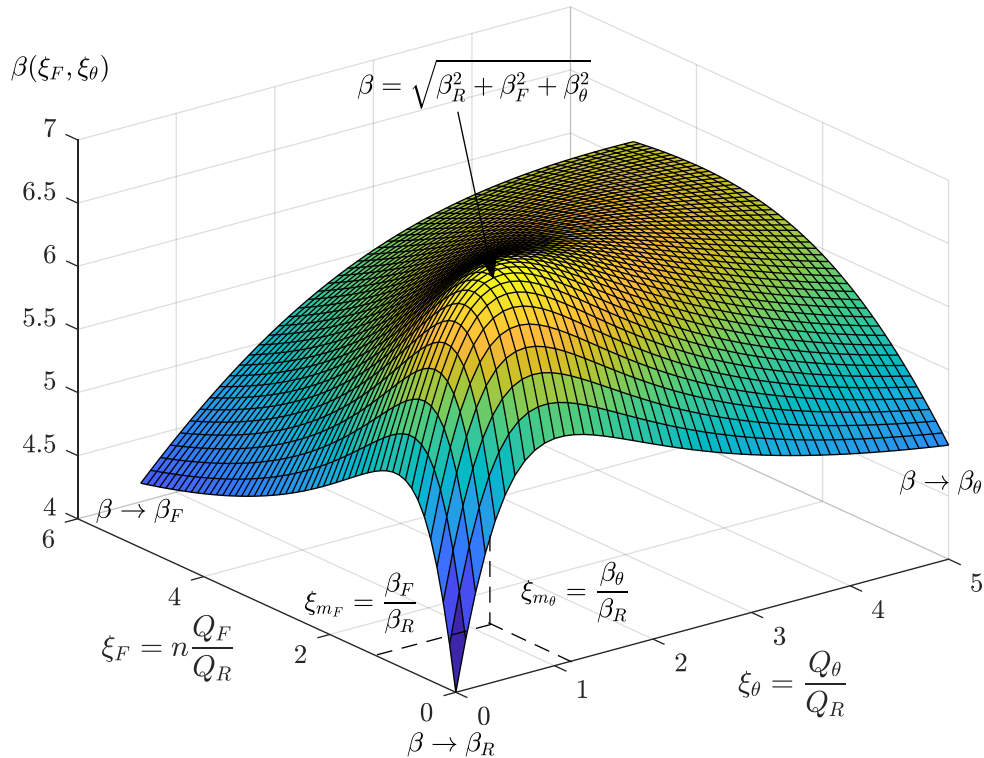

Figure 13  The influence of the model uncertainty on the RI of a nonlinear structural system with one action and one resistance, considering PSFs $\gamma_F; \gamma_R; \gamma_\theta$ more than 1.

## 4.5 The critical PSFs

**Definition 2**  *The critical PSF $\gamma_{c_i}$ of the variable $X_i$ is determined at the TRI $\beta_t$ assuming the variable $X_i$ dominates the system. It means the sensitivity of the variable $X_i$ is $\alpha_i = 1$ and thus the sensitivity of all other variables is negligible $\alpha_{j \neq i} = 0$.*

**Bakeer's theorem 2  The critical PSFs**

In a nonlinear structural system with variables $X_i, i = 1 \ldots N$, the RI is always bigger than the TRI $\beta_t$ if the following two conditions are satisfied:

(1) The PSFs are more than or equal to the critical PSFs $\gamma_{c_i}$ for each variable $X_i$, respectively.
(2) The PSFs are applied according to Definition 1.

The Bakeer's theorem 2 is a fairly straightforward result of Bakeer's theorem 1, considering the PRIs equal to the TRI, i.e. $\beta_i = \beta_t$. The critical PSFs for each variable can be determined by assuming that the variable is dominating the nonlinear system. The critical PSFs depend on the distribution and can be determined from Table 3, Table 4, Table 5, and Table 6 based on the percentile $p$ at which the characteristic value is calculated.

It is important to note that according to Definition 1, the critical PSFs of favourable actions must always be greater than or equal to 1, i.e. $\gamma_{F_c} \geq 1$. It differs from the EN 1990 format, where favourable actions have PSFs less than or equal to 1 and are introduced in a similar way to unfavourable actions.

Table 3   The critical PSFs for unfavourable actions according to the type of distribution. The characteristic values are calculated at $p$ percentile. Usually, $p$ is taken as 50% for the permanent actions, and 95%, or 98% for the variable actions.

| Lognormal distribution | $\gamma_{F_c} = \exp\left(Q_F(\beta_t - \Phi^{-1}(p))\right)$ |
|---|---|
| Normal distribution | $\gamma_{F_c} = \dfrac{1 + \beta_t V_F}{1 + \Phi^{-1}(p)V_F}$ |
| Gumbel distribution | $\gamma_{F_c} = \dfrac{1 - V_F \cdot \dfrac{\sqrt{6}}{\pi}(\gamma + \ln(-\ln(\Phi(\beta_t))))}{1 - V_F \cdot \dfrac{\sqrt{6}}{\pi}(\gamma + \ln(-\ln p))}$ |

Table 4   The critical PSFs for favourable actions according to the type of distribution. The characteristic values are calculated at $p$ percentile.

| Lognormal distribution | $\gamma_{F_c} = \exp\left(Q_F(\beta_t + \Phi^{-1}(p))\right)$ |
|---|---|
| Normal distribution | $\gamma_{F_c} = \dfrac{1 + \Phi^{-1}(p)V_F}{1 - \beta_t V_F}$ |

Table 5   The critical PSFs for favourable resistance or strength parameters according to the type of distribution. The characteristic values are calculated at $p$ percentile. Usually, $p$ is taken as 5%

| Lognormal distribution | $\gamma_{R_c} = \exp\left(Q_R(\beta_t + \Phi^{-1}(p))\right)$ |
|---|---|
| Normal distribution | $\gamma_{R_c} = \dfrac{1 + \Phi^{-1}(p)V_R}{1 - \beta_t V_R}$ |

*Table 6  The model critical PSFs according to the type of distribution. The characteristic values are calculated at $p$ percentile.*

| Lognormal distribution | $\gamma_{\theta_c} = \exp\left(Q_\theta(\beta_t - \Phi^{-1}(p))\right)$ |
|---|---|
| Normal distribution | $\gamma_{\theta_c} = \dfrac{1 + \beta_t V_\theta}{1 + \Phi^{-1}(p) V_\theta}$ |

## 5  Critical remarks on the OP-UP approach

The characterization of the system's behaviour as an OP or UP (Figure 2) can be better interpreted within the framework of the proposed theory of homogeneity. Considering a system with one unfavourable action, according to section 2.4, the OP or UP behaviour can be defined based on the DH $n_E$ as follows: the behaviour is OP if $n_E > 1$, and the behaviour is UP if $0 < n_E < 1$.

The purpose of this classification according to [32], is to provide simplified safe-side rules for the application of the PSF in nonlinear analysis. With exception of Basler's remarks about the beam example in Figure 1, the author found no proven scientific basis for this approach. However, the background might be explained for a nonlinear structural system $E = E(F)$ with one unfavourable action and one resistance, by considering the following two options:

- Option 1: the PSF $\gamma_F$ is applied to the action, i.e. $E_d = E(\gamma_F F_k)$. The resulting RI from this option is denoted by $\beta_1$;
- Option 2: the PSF $\gamma_F$ is applied to the effect, i.e. $E_d = \gamma_F E(F_k)$. It is also equivalent to applying the GSF $\gamma_R \gamma_F$ to resistance. The resulting RI from this option is denoted by $\beta_2$.

Based on the DH $n_E$, and by applying eq. (56) to each option, the following relations are obtained:

  (a) If the system is OP, i.e. $n_E > 1$, then $\beta_1 > \beta_2$.
  (b) If the system is UP, i.e. $0 < n_E < 1$, then $\beta_1 < \beta_2$.
  (c) If the system is linear homogenized, i.e. $n_E = 1$, then $\beta_1 = \beta_2$

The above conditions are correct, and they demonstrate a useful relative concept to compare the RI from option 1 with option 2 based on the nonlinear behaviour of the system (OP or UP). For general applications, however, this relative concept is insufficient to bring the TRI into a safe and economical range.

The application of $\gamma_F$ to the effect side or equivalently to the resistance side, has been seen in EN 1990 as compensation for the increased sensitivity of the system to the resistance in the UP case. However, the sensitivity of the action also increases in the OP system more than that for the linear system, and no compensation has been suggested for this option. Before discussing whether option 2 requires compensation or not, let's consider a third option, in which the GSF $\gamma_R \gamma_F$ is applied to action, and let's denote the resulting RI from option 3 by $\beta_3$.

Based on the DH $n_E$, and by applying eq. (56) to each option, the following relations can be obtained (Figure 14):

  (a) If the system is OP, i.e. $n_E > 1$, then $\beta_2 < \beta_1 < \beta_3$.
  (b) If the system is UP, i.e. $0 < n_E < 1$, then $\beta_3 < \beta_1 < \beta_2$.

(c) If the system is linear homogenized, i.e. $n_E = 1$, then $\beta_1 = \beta_2 = \beta_3$

Consequently, option 3 is safer than option 1 for the OP system. However, it is inaccurate to say here that $\gamma_R$ compensated for the increased sensitivity of the system to the action. It increases the safety of the OP system as can be seen from Figure 14, but the amount of increase is not correctly related to the TRI. The same applies to option 2 where $\gamma_F$ has been seen as compensation for the increased sensitivity of the system to the resistance. The application of $\gamma_F$ on the effect side increases the safety of the UP system but the amount is not justified in comparison with the TRI. According to Bakeer's theorem 2, the RI can be achieved for the UP system if $\gamma_R \geq \gamma_{R_c}$. However, this inequality is found to be satisfied by many material and resistance models (Table 7). Therefore, adding an increasing factor $\gamma_F$ to the resistance would only result in an over-safe and uneconomic design, which is not the purpose of using advanced computational nonlinear models.

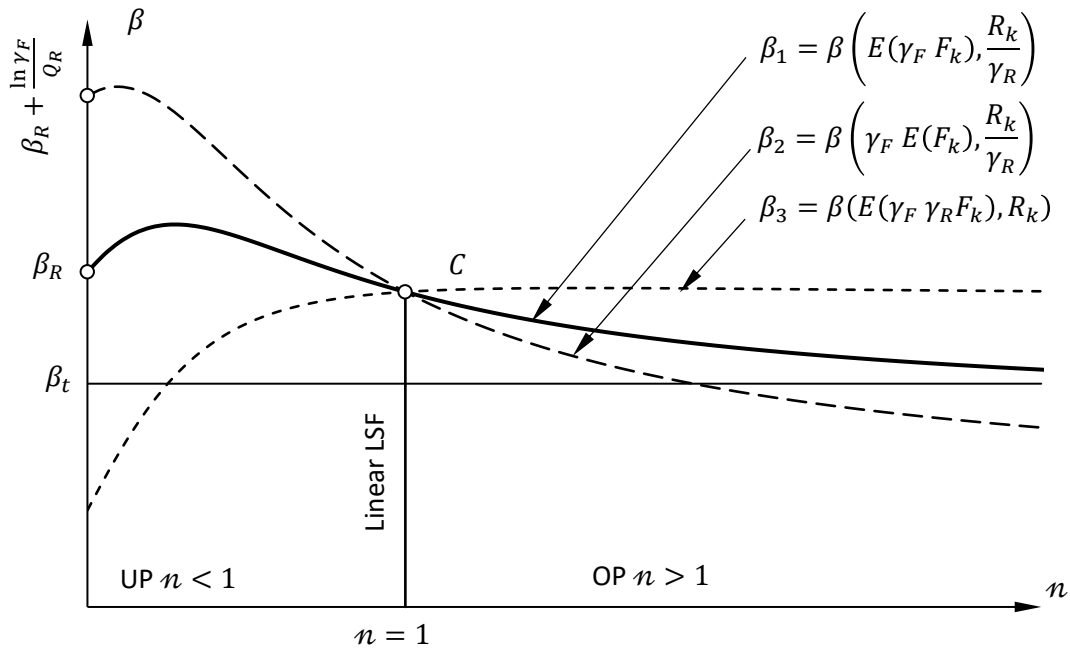

Figure 14    General demonstration of the relationship between the RI and the DH $n$. option 1: the PSF $\gamma_F$ is applied to action, option 2: the GSF $\gamma_F \gamma_R$ is applied to resistance. option 3: the GSF $\gamma_F \gamma_R$ is applied to action.

According to eq. (26) or eq. (5), $n_E$ can be determined by increasing all actions in the system with the same factor $\gamma$. This can also be learned from the example of Basler in Figure 1. However, increasing all actions with the same factor does not mean applying the coded PSFs, but rather finding the DH $n_E$ that characterizes the nonlinearity of the system as UP or OP. However, in the case of multiple actions, it is often that different PSFs need to be applied to different actions. As stated by eq. (38), the different contributions of actions to the effect are determined by the PDHs $n_{F_i}$ of each action and not $n_E$. Only in the case of a single action, we have $n_E = n_F$. Consequently, this may lead to incompatibility in the application of OP and UP classification in the case of multiple actions.

It has been explained in [32], how to use the PSFs according to the OP and UP classification (Figure 2). For a structural system with a single action and an UP behaviour, the PSF has been applied to the effect side. However, for a structural system with two actions and an UP behaviour, PSFs have been applied to the actions and the model PSF has been applied to the effect. The case with two actions is treated differently by [32] than the case with one action, even though both cases demonstrate an UP behaviour. The authors however stated that "*in practice, the situation may be more complex and more refined methods are needed*".

In a structural system with multiple actions, it is not always feasible to calculate the effect of a single action independently of the other actions. In many cases, this may lead to instabilities in the structural system, e.g. It is often impossible to determine the effect of single wind action in a membrane structure without the prestressing action. It is also not possible to determine the effect of a horizontal action in a masonry shear wall without the self-weight action (see the example in section 3.5). In both cases, the pre-stressing in membrane structures and the self-weight in masonry shear walls are essential for stabilizing the structural system. Without these stabilizing actions, it may not be possible to determine the effect of action according to structural mechanics. Thus, the classification of nonlinear systems as OP and UP does not lead to a simplified safe-sided rule, but rather to more complexity and incompatibility.

*Table 7    Comparison between the critical PSFs and the PSFs in Eurocodes for some material strength variables. The data for the stochastic variables are based on [73]*

| Stochastic variable | Dist. Type | COV | Mean | percentile % (Char value) | $\gamma_M$ | $\gamma_{M_c}$ |
|---|---|---|---|---|---|---|
| Steel yielding strength | Logn. | 0.05 | 1.00 | (0.83) | 1.00 | 1.00 |
| Concrete compression strength | Logn. | 0.1 | 1.00 | 5% | 1.50 | 1.18 |
| Rebar yielding strength | Logn. | 0.045 | 1.00 | 5% | 1.15 | 1.08 |
| Glulam bending strength | Logn. | 0.15 | 1.00 | 5% | 1.25 | 1.28 |
| Masonry compression strength | Logn. | 0.16 | 1.00 | 5% | 1.50 | 1.30 |
| Aluminium 0,2% limit | Logn. | 0.05 | 1.00 | (0.85) | 1.10 | 1.00 |
| Cone penetration test value | Logn. | 0.12 | 1.00 | 5% | 1.50 | 1.22 |
| Undrained shear strength | Logn. | 0.2 | 1.00 | 5% | 1.40 | 1.39 |

# 6    Solutions for engineering practice

The presented theory reveals profound implications for engineering practice and sets the basis for the application of PSFs in nonlinear structural systems. First, by introducing the homogeneity analysis of nonlinear systems as a tool for evaluating the impact of the basic variables on safety. Second, by determining the upper and lower bounds of the RI of a nonlinear structural system based on Bakeer's theorem 1. Third, by introducing the concept of nonlinearity-invariant critical PSFs based on Bakeer's theorem 2.

The results of this study provide a variety of options for assessing the safety of nonlinear structural systems. The practical implementations can, however, be classified into the following levels, from the simplest to the optimal, considering the trade-off between the aspects of simplicity and optimality:

## 6.1 Level 1: Critical PSF approach

It is the simplest approach and most suitable for engineering practice. The approach is based on a simple rule: *apply the nonlinearity-invariant critical PSF of each basic variable to the characteristic value according to Definition 1*. This approach does not require performing a homogeneity analysis to determine the DHs or the PDHs.

These nonlinearity-invariant critical PSFs can be calculated directly from Table 3, Table 4, Table 5, and Table 6 based on the statistical data about the actions, materials, and models, as well as the TRI of the code. The task of determining these nonlinearity-invariant critical PSFs is left to each code committee.

## 6.2 Level 2: Limited DH approach

Many nonlinear systems can have their critical PSFs reduced by providing some details about the range of variation of the DH or the RSP such as $\xi_R < \xi < \xi_F$. This may be explained by taking one action parameter and one resistance parameter as an example. Let's introduce the reduction factors $\kappa_F$ and $\kappa_R$ to the TRI. This reduction can be mapped to the critical PSFs by introducing reduction factors for the PRIs as follows: $\beta_F = \kappa_F \beta_t$ for the action and $\beta_R = \kappa_R \beta_t$ for the resistance.

These reduction factors $\kappa_R$ and $\kappa_F$ can be calculated by inserting the values of $\beta_F = \kappa_F \beta_t$ and $\beta_R = \kappa_R \beta_t$ in eqs. (58) and solving them for $\kappa_R$ and $\kappa_F$, simultaneously:

$$\kappa_R = \sqrt{\frac{\sqrt{1+\xi_F^2}\sqrt{1+\xi_R^2} - \xi_F \xi_R + 1}{\sqrt{1+\xi_F^2}\sqrt{1+\xi_R^2} + \xi_F \xi_R + 1}}; \quad \kappa_F = \kappa_R \frac{\xi_F\sqrt{1+\xi_R^2} + \xi_R\sqrt{1+\xi_F^2}}{\sqrt{1+\xi_R^2} + \sqrt{1+\xi_F^2}} \quad (59)$$

where $\kappa_F^2 + \kappa_R^2 \geq 1$.

The $\beta_t$ in Table 3 and Table 5 is allowed to be reduced by factors $\kappa_R$ for the resistance and $\kappa_F$ for the action if the RSP $\xi$ falls within the range $\xi_R < \xi < \xi_F$ or equivalently, if the DH $n$ falls within the range:

$$\frac{Q_R}{Q_F} \xi_R < n < \frac{Q_R}{Q_F} \xi_F \quad (60)$$

The values of $\kappa_R$ and $\kappa_F$ are given in Table 8 for several ranges of variations of RSP $\xi$.

This approach can be applied to the example of a cable element with lateral force in section 3.3. In this example, the DH varies in the range $2/3 \leq n < 1$, considering the COV of the action and resistance $V_F = 0.1$; $V_R = 0.05$, respectively, this gives $\xi_R = 1.33$; $\xi_F = 2$ the corresponding reduction factors can be calculated by eqs. (59) which give: $\kappa_R = 0.53$ and $\kappa_F = 0.85$. if the RI is $\beta_t = 3.8$ the critical PSFs can be determined at $\beta_F = 0.85 \cdot 3.8 = 3.23$ and $\beta_R = 0.53 \cdot 3.8 = 2.01$. Note that $\kappa_R$ and $\kappa_F$ are not sensitivity factors but are reduction factors of the lower bounds of RI, i.e. $\beta_R$ and $\beta_F$. These reduction factors are introduced because the structural system is not reaching the extreme cases of sensitivity.

Table 8     The reduction factors $\kappa_R$ and $\kappa_F$ for the range of variation $\xi_R < \xi < \xi_F$

| | | $\xi_F = 0$ | 0.2 | 0.4 | 0.6 | 0.8 | 1 | 2 | 4 | 6 | 8 | 10 | ∞ |
|---|---|---|---|---|---|---|---|---|---|---|---|---|---|
| $\xi_R = 0$ | $\kappa_R = 1$ | | 1.00 | 1.00 | 1.00 | 1.00 | 1.00 | 1.00 | 1.00 | 1.00 | 1.00 | 1.00 | 1.00 |
| | $\kappa_F = 0$ | | 0.10 | 0.19 | 0.28 | 0.35 | 0.41 | 0.62 | 0.78 | 0.85 | 0.88 | 0.90 | 1.00 |
| 0.2 | | | | 0.98 | 0.96 | 0.95 | 0.93 | 0.92 | 0.88 | 0.86 | 0.85 | 0.84 | 0.84 | 0.82 |
| | | | | 0.20 | 0.29 | 0.37 | 0.43 | 0.49 | 0.68 | 0.82 | 0.87 | 0.90 | 0.92 | 1.00 |
| 0.4 | | | | | 0.93 | 0.90 | 0.87 | 0.85 | 0.79 | 0.74 | 0.72 | 0.71 | 0.70 | 0.68 |
| | | | | | 0.37 | 0.45 | 0.51 | 0.56 | 0.72 | 0.85 | 0.89 | 0.92 | 0.93 | 1.00 |
| 0.6 | | | | | | 0.86 | 0.82 | 0.79 | 0.71 | 0.64 | 0.62 | 0.61 | 0.60 | 0.57 |
| | | | | | | 0.51 | 0.57 | 0.62 | 0.76 | 0.87 | 0.91 | 0.93 | 0.95 | 1.00 |
| 0.8 | | | | | | | 0.78 | 0.75 | 0.64 | 0.57 | 0.54 | 0.53 | 0.52 | 0.48 |
| | | | | | | | 0.62 | 0.67 | 0.80 | 0.89 | 0.92 | 0.94 | 0.95 | 1.00 |
| 1 | | | | | | | | 0.71 | 0.59 | 0.51 | 0.48 | 0.46 | 0.45 | 0.41 |
| | | | | | | | | 0.71 | 0.82 | 0.90 | 0.93 | 0.95 | 0.96 | 1.00 |
| 2 | | | | | | | | | 0.45 | 0.35 | 0.31 | 0.29 | 0.28 | 0.24 |
| | | | | | | | | | 0.89 | 0.94 | 0.96 | 0.97 | 0.98 | 1.00 |
| 4 | | | | | | | | | | 0.24 | 0.20 | 0.18 | 0.17 | 0.12 |
| | | | | | | | | | | 0.97 | 0.98 | 0.98 | 0.99 | 1.00 |
| 6 | | | | | | | | | | | 0.16 | 0.14 | 0.13 | 0.08 |
| | | | | | | | | | | | 0.99 | 0.99 | 0.99 | 1.00 |
| 8 | | | | | | | | | | | | 0.12 | 0.11 | 0.06 |
| | | | | | | | | | | | | 0.99 | 0.99 | 1.00 |
| 10 | | | | | | | | | | | | | 0.10 | 0.05 |
| | | | | | | | | | | | | | 1.00 | 1.00 |

## 6.3 Level 3: Homogeneity analysis approach

The homogeneity analysis in section 3 demonstrates a highly effective approach for applying the PSFs to complex nonlinear systems safely and economically. As the homogeneity analysis is well suited to computer implementation, it can be integrated with the structural analysis to assess the safety of the structure based on the TRI. A separate paper may address the various options and solutions that can be applied to this approach.

# 7 Conclusions

A novel and general theory is introduced in this paper to provide the necessary theoretical basis for applying the PSF method to nonlinear structural systems. It establishes, for the first time since the development of limit-state theory, the necessary key relationship between the PSF concept and the reliability theory of nonlinear structural systems. The following main conclusions can be drawn from the proposed theory:

– The theory of homogeneity has been able to provide a set of advantages and straightforward solutions for the application of PSFs to nonlinear structural systems.

- Whenever the PSFs are applied to a nonlinear structural system for a specific design or verification case, the RI always has an upper and lower bound (according to Bakeer's theorem 1). The upper and lower bounds are based solely on the statistical properties of the basic variables and are nonlinearity invariants.
- In a nonlinear structural system, the application of PSFs more than the critical PSFs for all basic variables guarantees that the RI of the system is more than the TRI (according to Bakeer's theorem 2). This provides a nonlinearity-invariant solution to apply the PSFs in engineering practice.
- In complex nonlinear structural systems, homogeneity analysis offers a highly effective approach for examining different design and verification cases.
- The incompatibilities associated with the OP-UP approach in EN 1990 have been addressed. It is advised to remove the provisions related to the over-/under linear behaviour from the code and replace them with the suggested solutions in section 6.1.
- For engineering practice, several solutions have been proposed based on the levels of simplicity and optimality, including the critical PSF approach, limited DH approach, and homogeneity analysis approach.
- It is recommended to apply the PSF to the corresponding characteristic value of the basic variable.

## Acknowledgement

The present theorems have been developed as a part of the author's activities on revisions of the Eurocodes, especially in the subcommittees on *prEN1990: Basics of structural and geotechnical design*, working group 3 of *CEN/TC 250/SC 10: Safety format and nonlinear problems/FEM applications* and *CEN-TC250-SC 10, Ad Hoc Group on Reliability background of the Eurocodes*.

It is with great gratitude that the author wishes to acknowledge *Prof. Dr.-Ing. Wolfram Jäger*, the convenor of working group 3, for the generous support and valuable discussions he has provided.